\RequirePackage{fix-cm}
\documentclass[smallextended]{svjour3}       
\smartqed  
\usepackage{graphicx}

\usepackage{subfigure}
\usepackage{booktabs}
\usepackage{url}
\usepackage{color}
\usepackage{breakcites}

\begin{document}

\title{Validating Gravity-Based Market Share Models Using Large-Scale Transactional Data
}
\subtitle{}


\author{Yoshihiko Suhara \and
        Mohsen Bahrami \and
        Bur\c{c}in Bozkaya \and
        Alex `Sandy' Pentland
}


\institute{Y. Suhara, M. Bahrami, A. Pentland \at
              The Media Lab, Massachusetts Institute of Technology, Cambridge, MA, USA \\
              \email{suhara@mit.edu}           
           \and
           M. Bahrami, B. Bozkaya \at
              Sabanci University, Istanbul, Turkey
}

\date{}

\maketitle

\begin{abstract}
Customer patronage behavior has been widely studied in market share modeling contexts, which is an essential step towards modeling and solving competitive facility location problems. Existing studies have conducted surveys to estimate merchants' market share and factors of attractiveness to use in various proposed mathematical models. Recent trends in Big Data analysis allow us to better understand human behavior and decision making, potentially leading to location models with more realistic assumptions. In this paper, we propose a novel approach for validating Huff gravity market share model, using a large-scale transactional dataset that describes customer patronage behavior in a regional scale. Although the Huff model has been well-studied and widely used in the context of competitive facility location and demand allocation, this paper is the first in validating the Huff model with a real dataset. Our approach helps to easily apply the model in different regions and with different merchant categories. Experimental results show that the Huff model fits well when modeling customer shopping behavior for a number of shopping categories including grocery stores, clothing stores, gas stations, and restaurants. We also conduct regression analysis to show that certain features such as gender diversity and marital status diversity lead to stronger validation of the Huff model. We believe we provide strong evidence, with the help of real world data, that gravity-based market share models are viable assumptions for competitive facility location models.
\keywords{Market share \and Huff model \and Customer patronage behavior \and Big Data analysis}
\end{abstract}

\section{Introduction}
During past decades and especially by the advent of the new machinery and technologies, the number of companies has been increasing dramatically, which has led to a highly competitive business environment. For example, in the UK there has been a sustained growth in total business population with a 64\% growth rate since 2000, and the number of companies has continuously increased during recent years. In 2016, it has increased by 197,000, which is equal to 4\% growth \footnote{\url{http://www.fsb.org.uk/media-centre/small-business-statistics/}}. To compete in such an environment, perhaps the biggest challenge for companies is to "optimally" locate their facilities to capture more demand and market share, while trying to alleviate the burden of their fixed and operational costs. This makes facility location decisions of critical importance to companies as such decisions must take into account the market environment to operate in and consumers' preferences.

For decades, companies have been trying to understand how customers are attracted to retail businesses so as to make effective decisions about where to open a new store of their chain. To address this challenge, a vast literature on facility location models has emerged. Many facility location models focus on locating retail stores in a competitive environment. An overall aim of such models is to maximize the market share captured as a result of the new locations opened, and consequently to maximize the profitability of the company for its shareholders \cite{Jensen:2001bo}\cite{Wernerfelt:1986bh}. Hence, decision makers must understand and model the underlying processes for retail patronage before the facility location models can be solved effectively.

Among various models of competitive facility location that are developed and available in the literature~ \cite{Berman:2009kk}\cite{Eiselt:1993ko}\cite{Plastria:2001wk}\cite{Drezner:1995ti}, we are especially interested in those with underlying market share models that consider customer shopping behavior and retail patronization. The main goal of patronization models is to derive a realistic estimate of how and where people shop, and consequently a retail facility’s market share. These models assume that the patronage behavior is influenced by multiple factors such as the retail facility’s attractiveness to customers, distance from customers’ location, and customers' purchasing power \cite{Drezner:2006dr}. Among various market share estimation approaches proposed, five main ones include proximity \cite{Hotelling:1929co}, deterministic utility \cite{Drezner:1994fo}, random utility~\cite{Leonardi:1984eg}~\cite{Drezner:1996im}, cover-based \cite{Drezner:2010kc}, and gravity-based \cite{Huff:1964ix} approaches. 

The first and the simplest approach is the proximity approach, which only considers the distance factor. Hotelling was the first to propose and use this model \cite{Hotelling:1929co}. Based on this model, a customer is more likely to patronize the facility closer to his or her location. The second approach is the deterministic utility approach introduced by Drezner \cite{Drezner:1994fo}, which suggests that customers are attracted differently to retail facilities. Therefore, proximity only is not sufficient anymore and a utility value is defined for allocation of the customers to the facilities. Yet, customers are assumed to spend most at the facility that is most attractive to them. The third kind of model was introduced to address the problem of “all or nothing” in deterministic utility models. The random utility model is an extension of the deterministic utility model, where the utility of the customer is drawn from a multivariate normal distribution of utility function \cite{Leonardi:1984eg}\cite{Drezner:1996im}. The fourth is the cover-based approach, where for each facility an influence circle with a certain radius is defined based on its attractiveness. Customers inside the circle are fully attracted by the facility in the center and those outside the influence circles of all the facilities are considered as `lost sales' \cite{Drezner:2010kc}.

The fifth and the most extensively used approach is the gravity-based approach. Estimating market share based on this approach was first introduced by Reilly \cite{Reilly:1931wa} and further extended by Huff \cite{Huff:1964ix}\cite{Huff:1966}. The Huff model approximates the probability of a customer’s patronization of a particular retail facility based on two factors: attractiveness and distance. This means that the more attractive shops (based on various relevant criteria) draw more customers, and people tend to visit shops closer to where they live or work. It is common in Huff-based models to approximate the market share of each facility based on the total number of visits or the total money customers spend, which translate into the calculated probabilities of patronizing each facility. Nakanishi and Cooper \cite{Nakanishi:1974ik} further propose an improved Huff model by developing a multiplicative competitive interaction (MCI) model that combines multiple dimensions of attractiveness into a single measure. Many extensions of the Huff model proposed by other researchers using different attractiveness factors and distance decay functions are also proposed \cite{Bozkaya:2010ca}\cite{Berman:2002vl}\cite{Aboolian:2007jd}\cite{Hodgson:2007kx}.

To explore the nature of customer behavior and patronization choice, Drezner applies the Huff model as part of a behavioral analysis based on manually collected survey information \cite{Drezner:2006dr}. The survey uses a set of merchants in Orange County, California and tracks subjects to analyze how and why these customers patronize these merchants. Her metric for verifying the estimated attractiveness levels derived from the survey data is the correlation between the theoretical model's results and the estimation based on the survey, where a high correlation was reported.

In this study, we take an approach similar to Drezner’s to model customer retail patronization, but this time relying on real transaction data collected from tens of thousands of customers’ credit card activities. The recent rise of Big Data analysis has led to many similar studies trying to model and understand urban-scale human behavior based on call records \cite{Isaacman:2011fe}\cite{Blondel:2012wt}\cite{Bogomolov:2014ey}, credit card transactions \cite{Singh:2015:PLoSONE}\cite{Hasan:2012bf}, GPS traces \cite{Zheng:2013ip}, etc. The Huff model is a very popular model used both in research, appearing in 27 out of 55 articles on competitive facility location modeling as reported by \cite{Drezner:2014bl}, and in business/retail applications \cite{Huff:2008wq}, yet to the best of our knowledge, there is no research on using real transactional data sets to test or validate Huff or similar gravity-based models. In contrast to the previous studies based on survey data for understanding shopping behavior, this paper presents a novel data-driven approach for patronage behavior analysis based on real-world transaction records. We believe such an approach also alleviates the limitations of survey-based studies related to data collection and data quality, by using readily available data that reflect real human behavior as opposed to drawing conclusions based on stated preferences of consumers.

The contribution and advantages of our approach include the following:
\begin{itemize}
    \item Competitive facility location models can now benefit from the validated use of Huff or similar gravity based models for better representation of reality in retail patronization and market share estimation.
    \item One can consider using the Huff model for distinct merchant categories to compare its performance across various categories.
    \item Our analysis reveals that the presence or lack of certain demographic features, such as gender diversity or marital status diversity, lead to better validation of the Huff model.
    \item Merchants and business owners can implement our validation approach in different geographical regions with different settings so that retail location decisions can be used with higher reliability.
    \item It is computationally inexpensive to perform our validation approach on a different transactional data set. This eliminates the need and associated costs to conduct surveys for data collection under different settings.
\end{itemize}

The rest of our paper is organized as follows. In Section 2, we present our validation methodology. Next we present the results of our validation using the transactional dataset. Finally, we provide concluding remarks and directions for future research.


\section{Methodology}

\subsection{The Huff Model}
The Huff model \cite{Huff:1964ix} is an economic model for estimating market share in relation to customer retail patronization decisions. This model is based on gravity models \cite{Reilly:1931wa}\cite{Anderson:QLkYDpnx}, which describe the magnitude of interaction by two factors, namely mass and distance. The Huff model uses merchant {\it attractiveness} for the mass factor and the {\it distance} between a customer and a merchant for the distance factor. The utility of customer $i$ visiting merchant $j$ can be formally defined as follows:
\begin{equation}
H_{ij} = \frac{A_{j}^\alpha}{D_{ij}^\beta},
\end{equation}
where $A_{j}$ is the attractiveness of merchant $j$, $D_{ij}$ is the distance between customer $i$ and facility $j$, and the parameters $\alpha$ and $\beta$ are used to adjust the sensitivity of the model to the two factors. To obtain the probability of customer $i$ visiting merchant $j$, $H_{ij}$ is normalized by the sum of all utility values for possible visits:
\begin{equation}
P_{ij} = \frac{H_{ij}}{\sum_{j' \in c(i)} H_{ij'}},
\end{equation}
where $c(i)$ denotes the set of merchants that customer $i$ could potentially visit.

\subsection{Data}
In this study, a large amount of credit card transaction records were used for the designed experiment. The dataset was collected from a major bank in a major city of an OECD country between July 2014 and June 2015. The transaction records contain customer IDs, spending amounts, merchant IDs, and their merchant categories. We filtered customers who have at least 10 transactions in the dataset. The statistics of the dataset are shown in Table \ref{table:dataset_summary}.

\begin{table}[ht]\label{table:dataset_summary}
\caption{Dataset summary.}
\begin{tabular}{c|c}\hline
Period & July 1, 2014 to June 30, 2015 \\
\# of transactions & 4,254,652 \\
\# of customers & 62,392 \\
Avg. transactions / customer & 68.19 \\\hline
\end{tabular}
\end{table}

\noindent
{\bf Customer information.} For each customer in the dataset, we have the following demographic information:
\begin{itemize}
 \item age;
 \item gender;
 \item marital status;
 \item education level;
 \item work status (employed by private sector, self-employed, etc.);
 \item income (as estimated by the bank);
 \item home location;
 \item work location.
\end{itemize}

\noindent
{\bf Merchant Category.} Table \ref{table:merchant_category_stat} shows the selected merchant categories and their corresponding number of transactions and descriptions. We have chosen these categories to compare patronage behavior over different types of merchants. Customers tend to visit grocery stores more often than other categories. During our experiment, we evaluated the consistency and inconsistency between these categories.

\begin{table}[ht]
\caption{Basic statistics of the top-4 most frequented merchant categories.}
\label{table:merchant_category_stat}
\begin{tabular}{c|c}\hline
Merchant category & \# of transactions \\\hline\hline
Grocery store (Grocery) & 1,089,614 \\
Gas station (GS) & 482,178 \\
Clothing store (Clothing) & 437,760 \\
Restaurants & 185,595 \\\hline
\end{tabular}
\end{table}
 
\subsection{Experimental Setting}
We split the dataset into 17 regions based on the administrative districts of the city of interest. The Huff model was fitted to each region for each merchant category. Therefore, we created 68 (17 regions x 4 merchant categories) Huff models for the experiment. For each region, we had a set of merchants belonging to the corresponding categories and a set of customers who visited these merchants. This resulted in the creation of a visit-count matrix $V_{ij}^{(r, c)}$, which consists of the visit count of customer $i$ to merchant $j$ that belongs to merchant category $c$ and is located in region $r$.

\noindent
{\bf Revenue Estimation for Attractiveness.} We use the total revenue of merchant $j$ in the dataset as the magnitude of its overall economic presence, and hence as its attractiveness measure ($A_{j}$.) Although total transaction count of merchant $j$ is an alternative option, the total revenue is more appropriate from the facility location perspective. In other words, a company tries to choose the right location for a new facility in order to maximize the profitability. As a result, the revenue information well represents the profitability and attractiveness of a merchant. We aggregate the transaction amount of a merchant by all customers as an approximated revenue of each merchant.

\noindent
{\bf Parameter Estimation.} Parameters $\alpha$ and $\beta$ are optimized to maximize the evaluation metric through Particle Swarm Optimization (PSO) technique \cite{Kenndy:1995th} within the range of $\alpha, \beta \in [0, 100]$. We also tested the OLS method \cite{Nakanishi:1974ik}, which is commonly used for the Huff model parameter estimation. However, the method significantly under-performed the PSO method. Therefore, we used PSO for parameter estimation in this experiment. To the best of our knowledge, this paper is the first to use PSO or any kind of derivative-free optimization technique to optimize the $\alpha$ and $\beta$ parameters of the Huff model.

\noindent
{\bf Evaluation metric.} We use Pearson's correlation between the estimated visit distribution calculated by the Huff model and the actual transaction-based visit distribution of a region as an evaluation method. The fitted model outputs the probability of a customer visiting a merchant, resulting in a visiting probability matrix $P_{ij}^{(r,c)}$ whose $(i, j)$-element is the probability of customer $i$ visiting merchant $j$ of category $c$ in region $r$.

We also conduct regression analysis to find significant indicators of the computational model performance. In this part of our analysis, we compare the following indicators and conduct regression analysis to verify their contributions to the computational model performance.

\noindent
{\bf Mobility diversity.}  We define the mobility diversity of a district $i$ as the entropy value of visited districts for shopping. That is, for a given district $i$, we aggregate the transactions of all customers in that district by all districts in the region where the customers purchased items. A higher entropy value indicates that customers living in a district visit diverse areas for shopping.

\noindent
{\bf Demographic diversity.}  For demographic diversity, we use {\it gender}, {\it marital status}, {\it education level}, and {\it job status} attributes of customers living in a district. For each district, we aggregate the demographic attribute counts to calculate the diversity of each attribute. We use entropy as a diversity metric. 

\noindent
{\bf Merchant diversity.} We calculate the entropy value of merchant category distribution for each district. If a district has exactly the same number of merchants for each merchant category, the entropy takes the highest value. We prepared this merchant diversity metric following the intuition that a skewed distribution of merchant categories possibly affects patronage behavior in a district or region.

\noindent
{\bf Merchant share bias.} We calculate merchant share bias based on the market share of the top-5 merchant shares in a district. We calculate the total transaction amount of merchants for each district and then we divide the total amount of the top-5 merchants by the total amount of all transactions in a district.

\noindent
{\bf Income inequality.} Based on the income information reported to the bank, we calculated the Gini coefficient of income distribution for each district for income inequality. As some customers reported their income as 0, we filter them out in order to get a reliable analysis.

We consider the indicators described above as independent variables and the Huff model performance value as a dependent variable.
We concatenated all district results to create a dataset with 68 (17 regions x 4 merchant categories) samples for the regression analysis.
All the variables were standardized by converting into z-scores for easier interpretation.

\section{Results and Discussion}

\subsection{Model Performance}
\begin{figure}[h]
  \centering
  \includegraphics[width=0.7\textwidth]{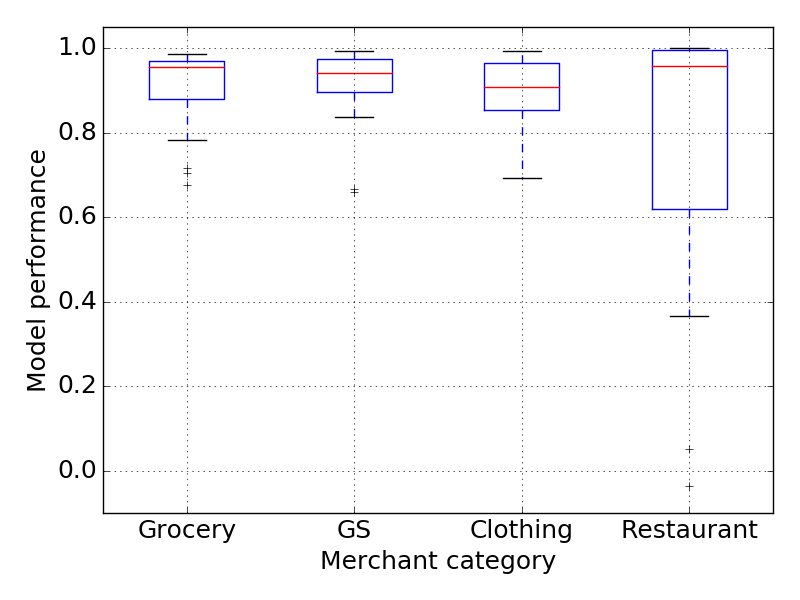}
  \caption{{\bf Distribution of model performance (Pearson's correlation) values for each merchant category.}}
\label{fig:eval_dist}
\end{figure}

Table \ref{table:model_performance} summarizes the basic statistics of the results and Figure \ref{fig:eval_dist} shows the boxplots of the model performance distributions for each merchant category. Detailed results including Pearson's correlation and the optimized parameter of each district are shown in the Appendix.

\begin{table}[h]
\centering
\caption{Huff model performance for each merchant category.}
\label{table:model_performance}
\begin{tabular}{ccccc}\hline\hline
Merchant category & Mean & Std & Max & Min \\\hline
Grocery & 0.8935 & 0.1068 & 0.9850 & 0.6753 \\
GS & 0.9050 & 0.1011 & 0.9928 & 0.6595 \\
Clothing & 0.8852 & 0.0930 & 0.9924 & 0.6916 \\
Restaurant & 0.7586 & 0.3418 & 1.0000 & -0.0370 \\\hline
\end{tabular}
\end{table}

 
The models in all categories perform well as their mean/median values of Pearson's correlation values exceed $0.7$ regardless of merchant categories. Except for the Restaurant category, the Pearson's correlation metric of the worst-performing district of each category is still above $0.65$. The results indicate that the Huff models based on transaction data robustly capture customer patronage behavior in these categories.

However, the Huff model in the Restaurant category has relatively unstable performance compared to the other categories. Four districts have less than $0.5$ Pearson correlation values and the worst performance shows $-0.037$. We consider that the main reason of the unstable performance of the Huff model in the Restaurant category arising from the fact that customers' patronage behaviors do not fully follow the Huff model's assumption. That means that people often choose to go to restaurants in distant locations with various attractiveness measures (other than the total revenue of the merchant) that are not captured in our model. One can view restaurant patronage as a more hedonic way of ``shopping" experience, where customers with a variety of tastes and expectations may choose to patronize a variety of places around the city to fulfill their expectations.

\begin{figure*}[ht]
  \centering
        	\subfigure[]              	{ \includegraphics[width=0.40\textwidth]{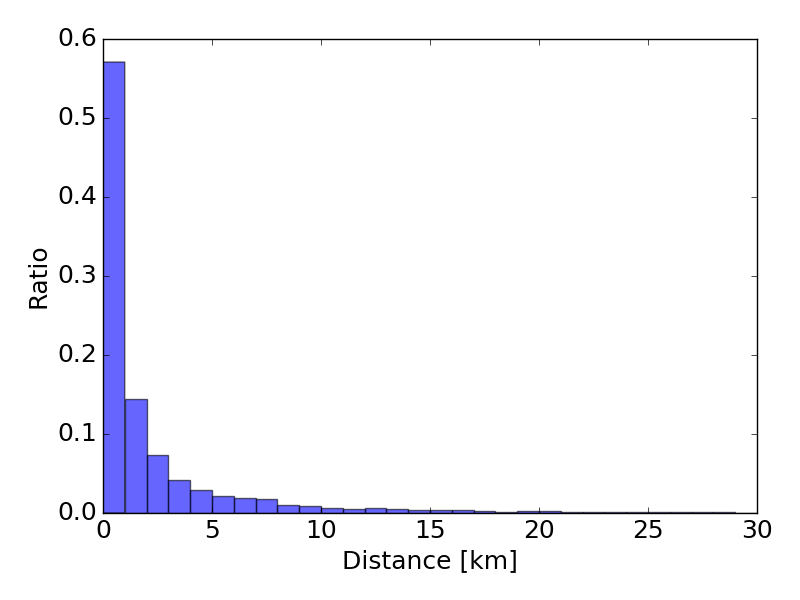} \label{fig:min_dist_5411_1}}
        	\subfigure[]              	{ \includegraphics[width=0.40\textwidth]{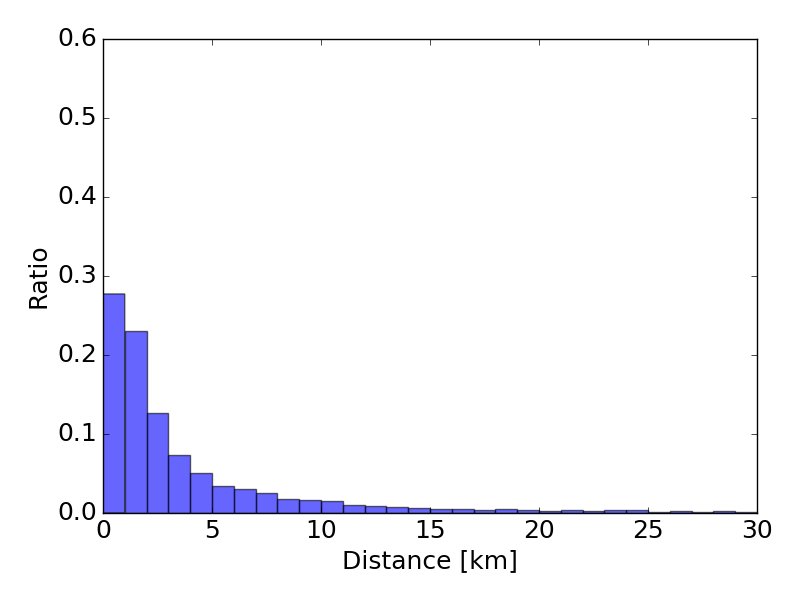} \label{fig:min_dist_5541_1}}
    	\vspace{-0.2cm}
        	\subfigure[]              	{ \includegraphics[width=0.40\textwidth]{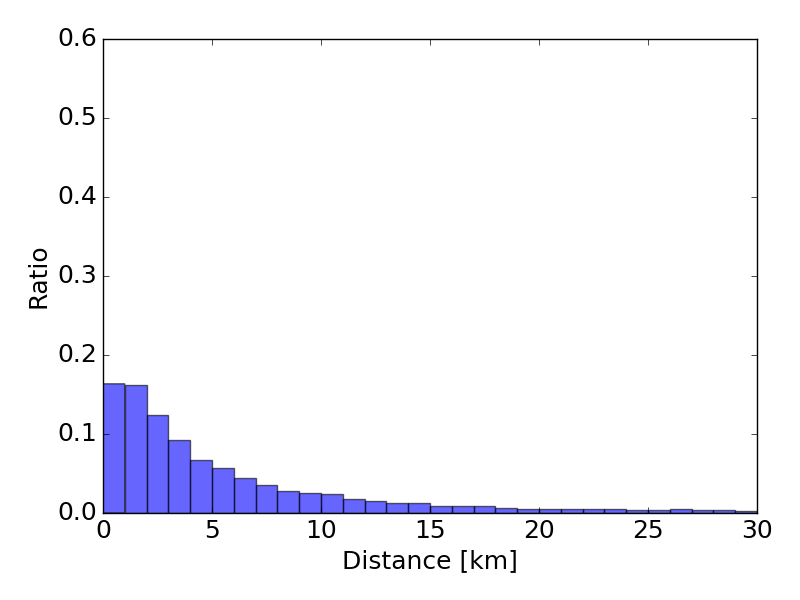} \label{fig:min_dist_5691_1}}
        	\subfigure[]              	{ \includegraphics[width=0.40\textwidth]{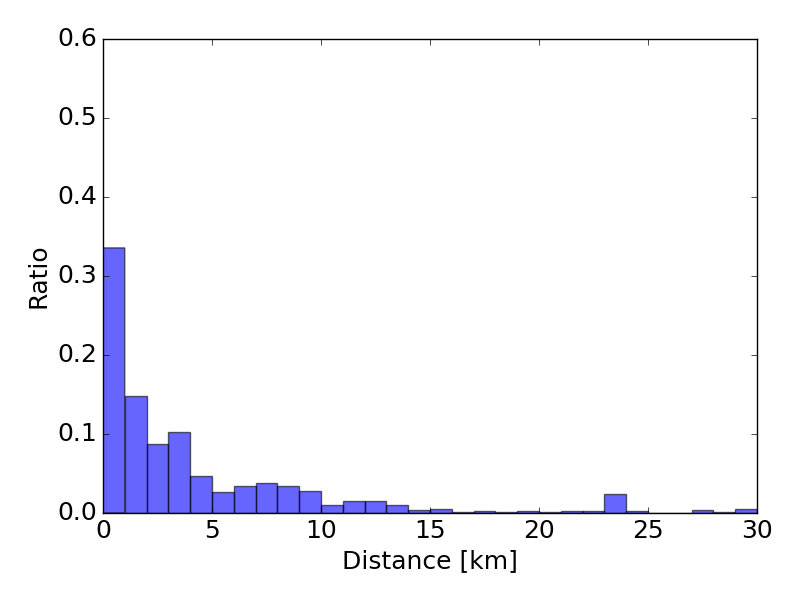} \label{fig:min_dist_5812_1}}
    	\vspace{-0.2cm}
  \caption{Distribution of the distance between visited merchants and customer's home/work locations (closer location is taken) of each merchant category: (a) Grocery, (b) GS, (c) Clothing, and (d) Restaurant}
  \label{fig:min_dist_dist}
\end{figure*}

To verify our interpretations, we analyze the distribution of the distance between visited merchants of these four categories and customers' home/work locations (closer location is taken.) The distributions are shown in Figure \ref{fig:min_dist_dist}. As shown in Figure \ref{fig:min_dist_5411_1}, \ref{fig:min_dist_5541_1}, and \ref{fig:min_dist_5691_1}, the visited merchants of the Grocery, GS and Clothing categories are basically located close to the customers' home/work locations while the distance distribution of restaurants contains long distance values as shown in Figure \ref{fig:min_dist_5812_1}. The results support our interpretation of the Huff model performance and also show a limitation of modeling patronage behavior with the Huff model based on transaction data. Despite this argument, we see that the model performance for the Restaurant category still suggests that the Huff model based on transactional data still performs reasonably well in several districts since the Huff model's performance in 13 out of 17 districts is higher than $0.5$.

The patronage behavior of gas stations is a great example of the Huff model being used on transaction data, among the four categories considered. We observe that the most frequented gas stations are in close proximity to the customers' home/work locations. Moreover, the mean value of the Huff model performance in the GS category is highest ($0.905$.) The result confirms the fact that customers often do stop by their popular gas stations in the vicinity of their home/work locations.

\subsection{Regression Analysis}
Table \ref{table:adjusted_r_squared} shows the adjusted ${\rm R}^2$ values of regression analysis for all four merchant categories. As shown in the table, the diversity measure indicates reasonably high Huff model performance for the Grocery and Clothing categories. On the other hand, the regression models do not perform well in establishing a link between diversity measures and Huff model performance for the GS and Restaurant categories. Furthermore, the regression models do not show any statistical significance in the $\beta$ coefficient values of the diversity indicators for these two categories.

\begin{table}[ht]
\caption{Adjusted ${\rm R}^2$ values of the regression analysis results.}
\label{table:adjusted_r_squared}
\begin{tabular}{c|c}\hline
Merchant category & Adjusted ${\rm R}^2$ score \\\hline\hline
Grocery & ${\bf 0.638}$ \\
GS & $-0.281$  \\
Clothing & ${\bf 0.557}$ \\
Restaurant & $0.150$  \\\hline
\end{tabular}
\end{table}

Table \ref{table:regression_analysis} shows the $\beta$ coefficient output of the regression models for (a) Grocery and (b) Clothing categories. The table summarizes the $\beta$ coefficient value of each indicator with 95\% confidence interval. We show the regression analysis results of the GS and Restaurant categories in the Appendix since we did not confirm any statistical significance in all the indicators for these categories.

The bold-faced values (i.e., statistically significant coefficients) in Table \ref{table:regression_analysis} suggest that gender diversity is positively correlated with the Huff model performance while marital status diversity is negatively correlated. A high gender diversity value means that males and females are equally distributed in a district. The gender diversity takes the highest value when the male/female ratio is one. In other words, a skewed distribution of male/female customers in a district makes the gravity model difficult to fit. On the other hand, marital status diversity is negatively correlated with the Huff model performance. The result follows our intuition that single and married customers have significantly different shopping styles. That is, the Huff model cannot simply generalize the patronage behavior in a district as the marital status diversity increases within the district. 


No significant statistical results can be seen in other indicators like mobility, merchant, and income diversity. Originally, we hypothesized that the mobility diversity and the merchant diversity would correlate with the Huff model performance. For instance, the mobility diversity is a direct indicator of the lifestyle of customers living in a district. Therefore, we would hypothesize that a high mobility diversity value of a district would make the Huff model difficult to fit.

Although the Huff model works well for the GS category, the regression analysis does not perform well in the GS category. Our interpretation of this is that the diversity features we propose are simply not indicative of the model performance across various districts in the region.


\begin{table}[ht]
  \centering
  \caption{OLS regression model between Huff model performance (i.e., the Pearson's correlation values) and indicators. $^*$, $^{**}$ denote  $p < 0.05, 0.01$ respectively. Bold face denotes the $\beta$ coefficient is statistically significant.}\label{table:regression_analysis}
  (a) Grocery \\
\begin{tabular}{lrr}\hline
Indicator                 & $\beta$ coefficient & Confidence interval (95\%)   \\ \hline\hline
Mobility diversity        & $-0.1799$           & $[-0.6516, 0.2917]$  \\
Merchant diversity        & $-0.2038$          & $[-0.7601, 0.3524]$  \\
Merchant monopoly         & $0.0586$           & $[-0.3650, 0.4822]$  \\
Gender diversity          & ${\bf 2.5007}^{**}$     & $[1.1776, 3.8239]$   \\
Marital status diversity  & ${\bf -2.4411}^{**}$    & $[-3.6434, -1.2388]$ \\
Education level diversity & $-0.3585$          & $[-0.8686, 0.1516]$ \\
Job status diversity      & ${\bf 0.5106}^{*}$      & $[0.0858, 0.9355]$  \\
Income inequality         & $0.2643$            & $[-0.2817, 0.8103]$  \\ \hline
\end{tabular}
\\
\vspace{3em}
  (b) Clothing \\
\begin{tabular}{lrr}\hline
Indicator                 & $\beta$ coefficient & Confidence interval (95\%)   \\ \hline\hline
Mobility diversity        & $0.1355$             & $[-0.3742, 0.6453]$  \\
Merchant diversity        & $0.5748$         & $[-0.0263, 1.1760]$  \\
Merchant monopoly         & $-0.4263$        & $[-0.8842, 0.0315]$  \\
Gender diversity          & ${\bf 1.4881}^{*}$        & $[0.0581, 2.9181]$   \\
Marital status diversity  & ${\bf -1.3081}^{*}$       & $[-2.6076, -0.0087]$ \\
Education level diversity & ${\bf -0.8321}^{**}$      & $[-1.3834, -0.2808]$ \\
Job status diversity      & $0.2080$             & $[-0.2511, 0.6672]$  \\
Income inequality         & $-0.3274$            & $[-0.9174, 0.2627]$  \\ \hline
\end{tabular}
\end{table}

\begin{figure*}[ht]
  \centering
        	\subfigure[]              	{ \includegraphics[width=0.40\textwidth]{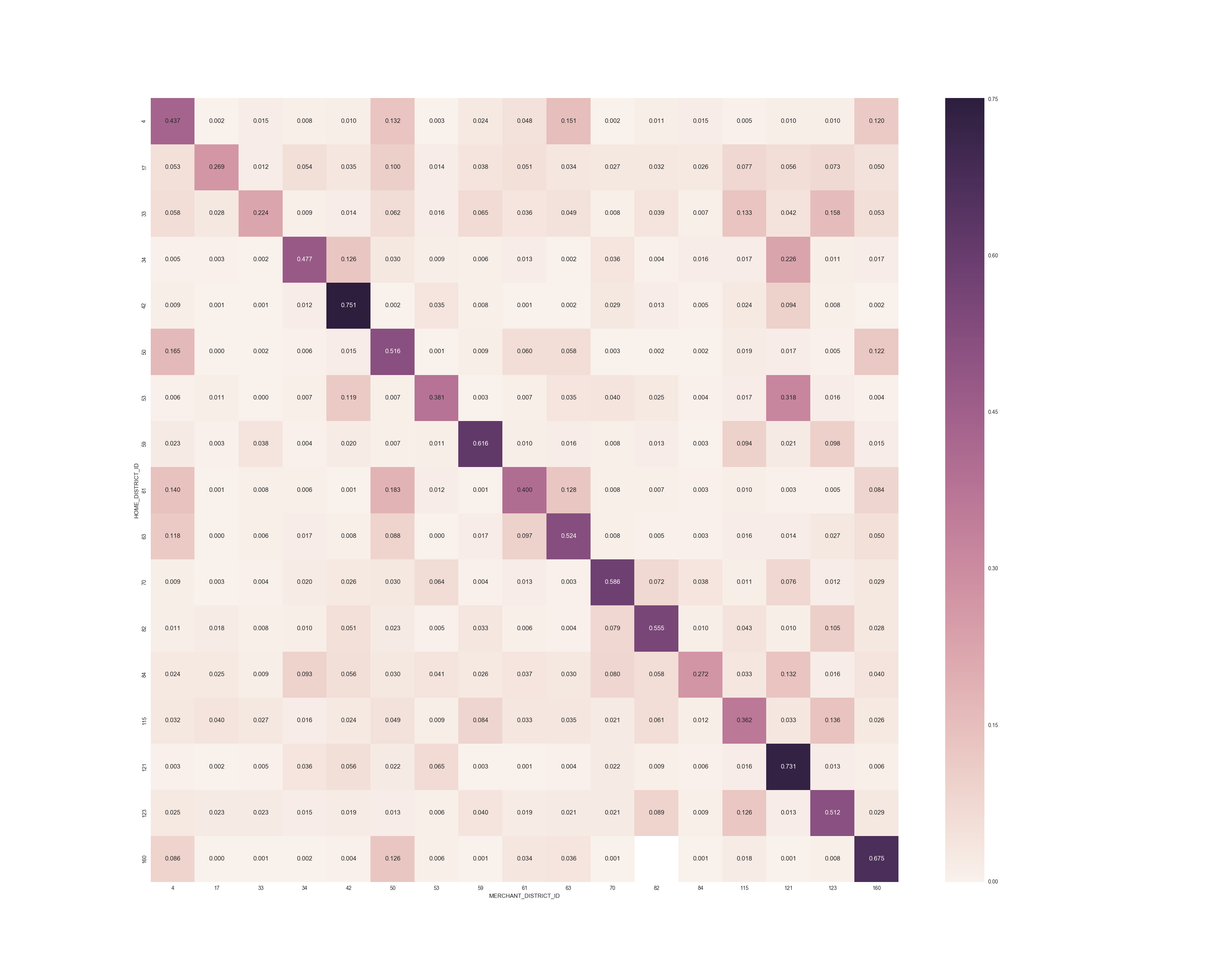} \label{fig:mobility_5411}}
        	\subfigure[]              	{ \includegraphics[width=0.40\textwidth]{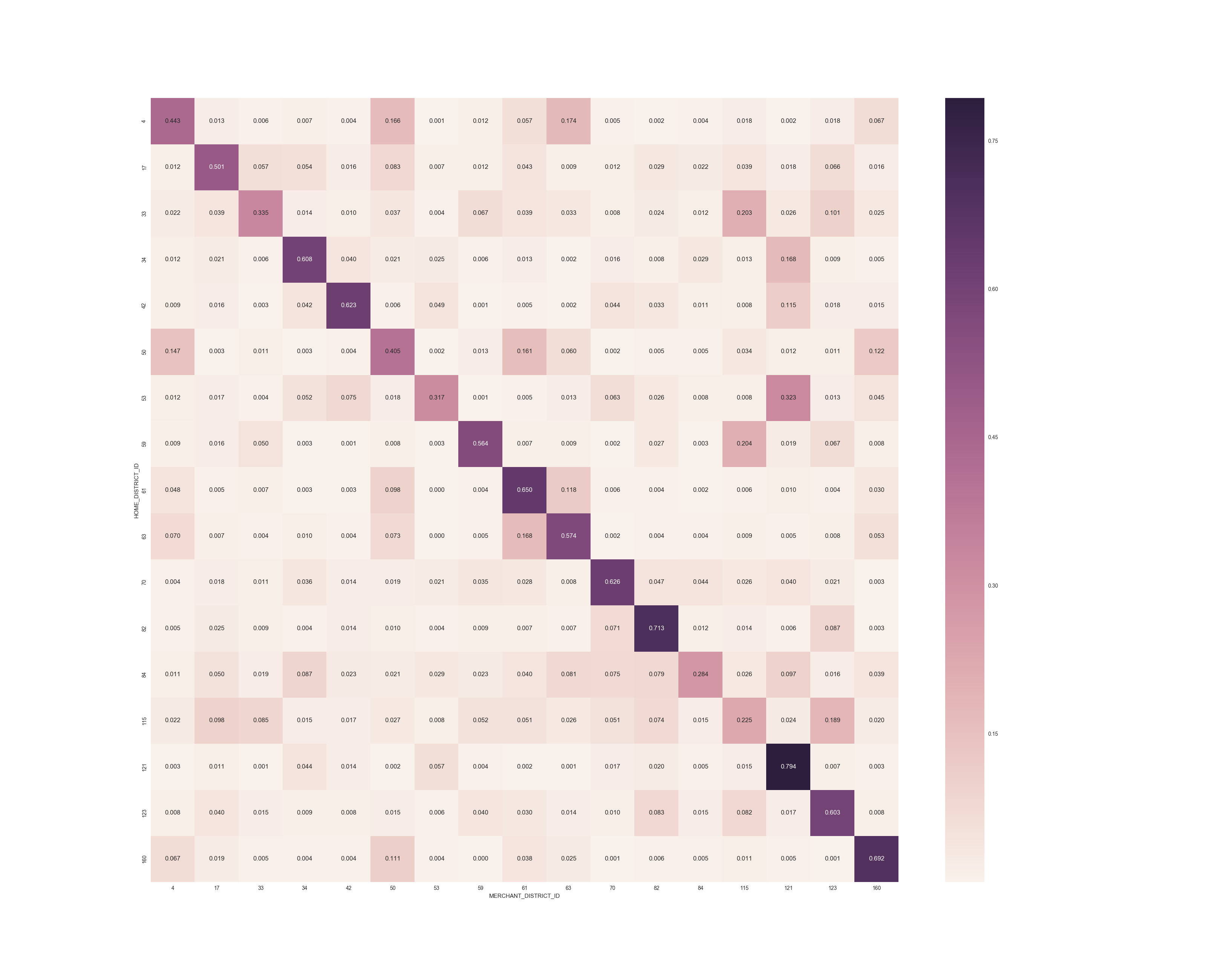} \label{fig:mobility_5541}}
    	\vspace{-0.2cm}
        	\subfigure[]              	{ \includegraphics[width=0.40\textwidth]{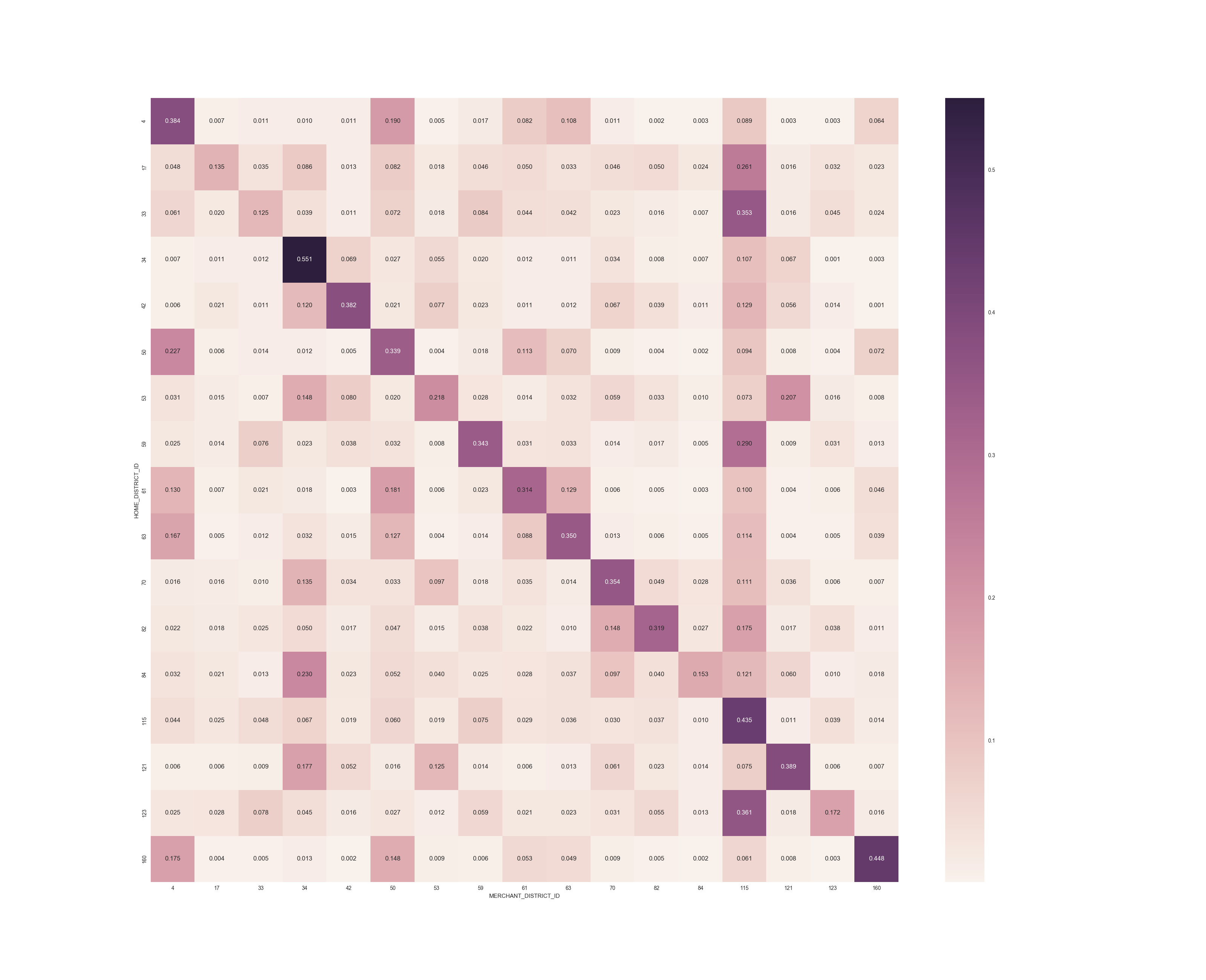} \label{fig:mobility_5691}}
        	\subfigure[]              	{ \includegraphics[width=0.40\textwidth]{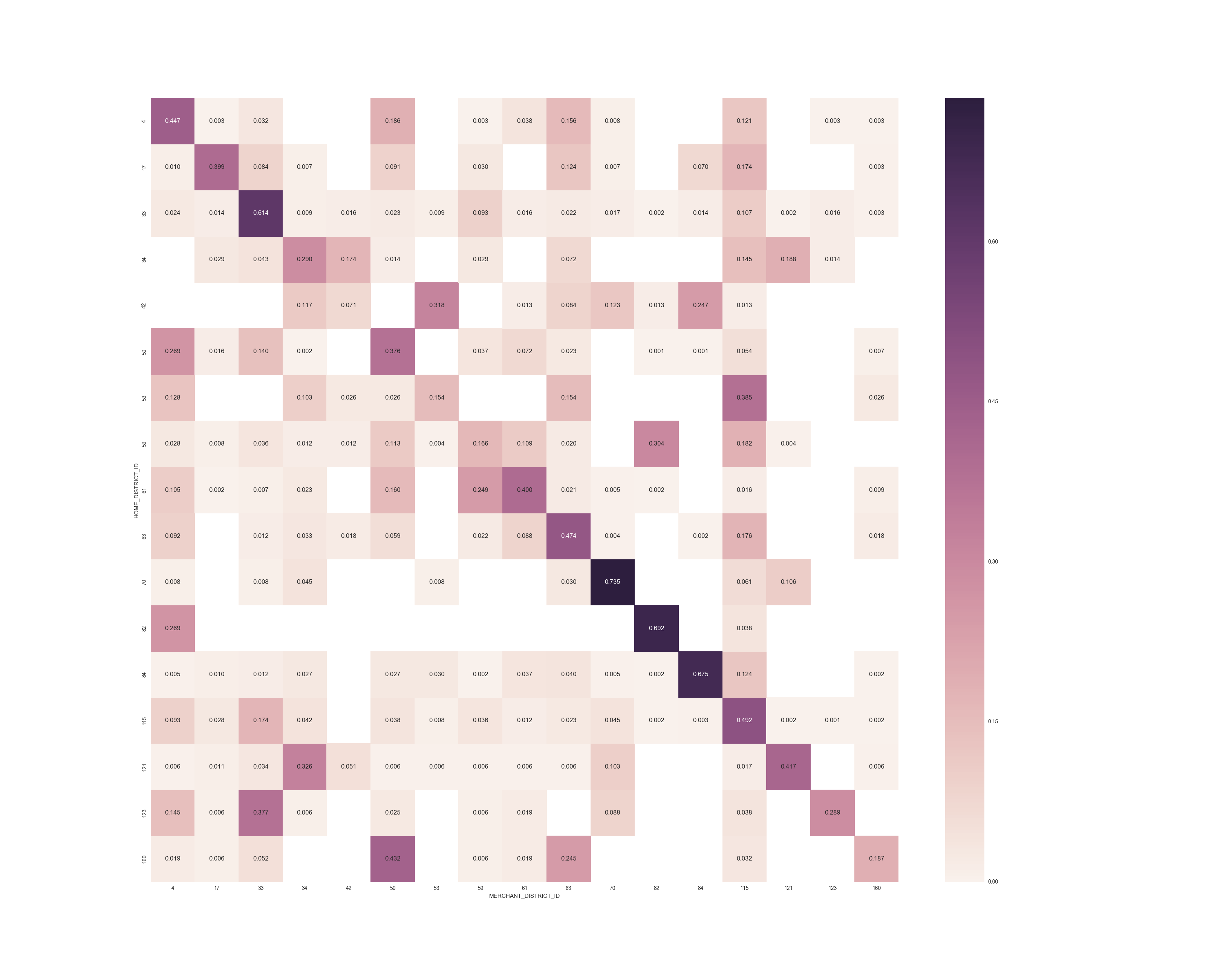} \label{fig:mobility_5812}}
  \caption{Mobility patterns of each merchant category: (a) Grocery, (b) GS, (c) Clothing, and (d) Restaurant.}
  \label{fig:mobility_patterns}
\end{figure*}

We have also conducted another type of mobility analysis to understand the differences regarding customers' shopping behavior for each merchant category. Fig. \ref{fig:mobility_patterns} shows the distributions of mobility patterns for each category. The $x$-axis and $y$-axis of these figures represent the district of a merchant and the district of a customer's home location, respectively. The numbers are normalized by row. For instance, the cell ($i$, $j$) is the normalized transaction frequency of merchants located in district $j$ by customers who live in district $i$. It is intuitive that the diagonal line basically has the highest values as customers mostly visit merchants in the same district as they live. However, Fig. \ref{fig:mobility_patterns} (d) shows that merchants in the Restaurant category have more biased distributed with respect to the mobility patterns.

\section{Summary and Conclusions}
In this paper, we have proposed a novel approach in validating a widely used customer retail patronage and market share estimation model, namely the gravity-based Huff model, using transactional data. Our approach applies the Huff model that consists of the attractiveness and distance factors to explain customer behavior. Our computational results have shown that the Huff model performs well in terms of the Pearson's correlation value calculated between the predicted market share and real market share.

Our study is the first to validate the Huff model with a large-scale transactional dataset to produce realistic representation of customer patronage behavior. Since gravity-based models such as Huff are widely used in competitive facility location optimization, our study provides key insight for reliable use of Huff or other gravity-based models in facility location optimization. Compared to the conventional survey-based approaches, the major advantages of our transaction-based approach include: (a) no requirement for surveys where data collection cost and data quality might be an issue, (b) the ability to directly compare different categories of shopping, and (c) ease of computational implementation in terms of computational complexity and time so that the model parameters can be updated in a periodic manner (e.g., daily, weekly, quarterly etc.) and also with different data sets.

As we have shown in our analysis, the performance of the Huff model varies between categories. For certain categories, additional criteria may need to be included in attractiveness calculations, or human behavior may simply be too complex to model using a gravity-based approach. However, we believe that our approach provides various benefits which cannot be obtained from the conventional (survey-based) approaches. On the other hand, we would like to emphasize that survey-based approaches can collect more fine-grained information and these two approaches can be complementary to one another. In this regard, combining survey-based information and transaction-based information to build a sophisticated shopping behavior model would be a future direction. Another possibility for future study might be validating and comparing the performance of other market share estimation models with the Huff model.

\section*{Acknowledgements}
The authors are grateful to the financial institution that provided the credit card transaction data for this research.

\section*{Supporting Information}

\begin{table}[ht]
  \centering
  \caption{Grocery}
  \begin{tabular}{lrrrrr}
\toprule
{} &  Avg. distance &   $\alpha$ &  $\beta$ & Pearson $r$ & $p$-value \\
districtid &            &            &             &           &                \\
\midrule
4          &   8.073920 &  44.092704 &   54.290707 &  0.957693 &   1.815968e-50 \\
17         &   7.193814 &  50.098677 &  100.000000 &  0.833162 &   6.116027e-14 \\
33         &   9.475784 &  25.459851 &  100.000000 &  0.450857 &   2.355955e-04 \\
34         &   8.946319 &   0.776089 &    0.026872 &  0.985043 &   5.379011e-46 \\
42         &   9.842346 &   0.869327 &    0.486844 &  0.968569 &   1.247863e-61 \\
50         &   8.526481 &   0.879940 &    0.000000 &  0.978794 &  3.661174e-134 \\
53         &  10.068181 &  44.430748 &   82.520949 &  0.906168 &   2.923771e-14 \\
59         &   9.293074 &   0.813904 &    0.000000 &  0.981842 &   3.553790e-55 \\
61         &   8.975816 &   0.727393 &    1.654360 &  0.884556 &   3.198309e-44 \\
63         &   8.391520 &  40.762860 &  100.000000 &  0.923845 &   3.488153e-63 \\
70         &   7.241697 &   3.612020 &   31.386009 &  0.632991 &   1.179105e-08 \\
82         &   6.566708 &   0.524779 &    1.277479 &  0.898500 &   5.617862e-30 \\
84         &   9.907927 &   0.581690 &    0.923136 &  0.715644 &   6.018784e-14 \\
115        &   8.304005 &  29.783773 &  100.000000 &  0.667452 &   2.556054e-17 \\
121        &   7.396679 &   0.858752 &    0.044139 &  0.965963 &   1.453555e-80 \\
123        &   6.220508 &   0.277481 &    1.423520 &  0.781789 &   1.194598e-22 \\
160        &   8.681894 &  34.661848 &   56.752011 &  0.953576 &   1.277798e-69 \\
\bottomrule
\end{tabular}
\end{table}

\begin{table}[ht]
  \centering
  \caption{Gas stations}
\begin{tabular}{lrrrrr}
\toprule
{} &  Avg. distance &   $\alpha$ &  $\beta$ & Pearson $r$ & $p$-value \\
districtid &            &            &             &           &               \\
\midrule
4          &  11.006358 &   0.592850 &    0.023663 &  0.940351 &  2.206558e-09 \\
17         &  10.139055 &   0.988860 &    1.424924 &  0.978365 &  1.124570e-11 \\
33         &  12.166465 &  40.288062 &  100.000000 &  0.978458 &  3.783059e-03 \\
34         &  12.719812 &   0.502571 &    0.851630 &  0.992763 &  1.772337e-11 \\
42         &  15.457192 &   0.983946 &    0.106719 &  0.974016 &  2.275157e-14 \\
50         &  11.539085 &   0.497945 &    0.310459 &  0.952412 &  2.336259e-13 \\
53         &  12.977294 &   0.698021 &    0.000000 &  0.867928 &  1.132502e-02 \\
59         &  11.643731 &  35.637848 &  100.000000 &  0.837396 &  2.501177e-03 \\
61         &  11.052237 &   0.711459 &    0.069953 &  0.896547 &  1.335805e-09 \\
63         &  10.527793 &  23.877870 &  100.000000 &  0.659504 &  4.554565e-04 \\
70         &   7.641760 &   0.507755 &    0.000000 &  0.930629 &  4.876992e-07 \\
82         &   9.296232 &  23.872853 &  100.000000 &  0.980732 &  7.087527e-10 \\
84         &  11.699089 &   0.698974 &   11.551928 &  0.966626 &  1.264309e-06 \\
115        &  11.732630 &   3.930043 &    2.705459 &  0.902851 &  5.778257e-05 \\
121        &  11.479532 &   0.453450 &    0.000000 &  0.895071 &  1.881786e-08 \\
123        &   8.481359 &   0.037286 &    0.478554 &  0.666772 &  9.202447e-03 \\
160        &  11.430354 &   0.748378 &    0.112061 &  0.964259 &  7.938458e-12 \\
\bottomrule
\end{tabular}
\end{table}

\begin{table}
  \centering
  \caption{Clothing}
\begin{tabular}{lrrrrr}
\toprule
{} &  Avg. distance &   $\alpha$ &  $\beta$ & Pearson $r$ & $p$-value \\
districtid &            &           &           &           &               \\
\midrule
4          &   9.158209 &  0.815865 &  1.191384 &  0.933204 &  3.323991e-63 \\
17         &  11.612848 &  1.129118 &  0.670492 &  0.970486 &  4.455518e-25 \\
33         &  11.131178 &  1.111029 &  2.365670 &  0.726212 &  1.762807e-16 \\
34         &  12.620926 &  1.014079 &  0.438777 &  0.854472 &  3.350070e-74 \\
42         &  17.460296 &  2.551279 &  3.436768 &  0.964759 &  2.092163e-39 \\
50         &  11.768580 &  0.799089 &  1.801190 &  0.811081 &  2.119634e-49 \\
53         &  13.393870 &  1.251408 &  3.273314 &  0.931657 &  1.266878e-27 \\
59         &  13.045309 &  0.863173 &  0.000000 &  0.691551 &  1.686600e-13 \\
61         &   9.828976 &  0.939626 &  0.000000 &  0.864188 &  1.346881e-37 \\
63         &   9.904244 &  0.701541 &  0.000000 &  0.908179 &  1.142956e-53 \\
70         &   9.030380 &  0.792165 &  0.918037 &  0.924041 &  6.768797e-26 \\
82         &   7.563128 &  1.052447 &  0.000000 &  0.977592 &  2.047455e-31 \\
84         &  11.273849 &  0.906613 &  2.291778 &  0.741179 &  4.978748e-10 \\
115        &  13.782368 &  1.065176 &  0.000000 &  0.878257 &  7.645000e-97 \\
115        &  13.782368 &  1.065176 &  0.000000 &  0.878257 &  7.645000e-97 \\
121        &  10.076457 &  1.356190 &  0.256458 &  0.992374 &  7.172498e-50 \\
123        &   8.647519 &  1.047006 &  0.000000 &  0.972279 &  3.175114e-26 \\
160        &   9.861500 &  0.826446 &  0.044827 &  0.907266 &  2.403986e-33 \\
\bottomrule
\end{tabular}
\end{table}

\begin{table}
  \centering
  \caption{Restaurants}
\begin{tabular}{lrrrrr}
\toprule
{} &  Avg. distance &   $\alpha$ &  $\beta$ & Pearson $r$ & $p$-value \\
districtid &            &               &             &           &               \\
\midrule
4          &   8.246086 &  1.694699e+00 &   14.005347 &  0.705706 &  3.513858e-04 \\
17         &  12.973318 &  3.609896e+01 &  100.000000 & -0.037036 &  9.306203e-01 \\
33         &  12.393137 &  2.284407e-01 &    2.121636 &  0.620474 &  1.584134e-03 \\
34         &  12.342009 &  1.000000e+02 &   46.690396 &  0.956793 &  1.102762e-06 \\
42         &  14.321486 &  0.000000e+00 &  100.000000 &  0.521633 &  4.783672e-01 \\
50         &  10.063284 &  4.376173e+01 &  100.000000 &  0.051864 &  7.932417e-01 \\
53         &  11.459094 &  7.655047e+01 &    0.403219 &  1.000000 &  0.000000e+00 \\
59         &  10.507480 &  1.200341e+00 &   11.290621 &  0.906397 &  1.909024e-03 \\
61         &   9.491864 &  8.194770e+01 &   48.867638 &  0.957679 &  1.333397e-05 \\
63         &  10.540970 &  3.694519e+01 &  100.000000 &  0.903863 &  1.347535e-04 \\
70         &  11.275803 &  1.482044e+00 &    0.415506 &  0.996231 &  2.128370e-05 \\
82         &  15.038343 &  6.727115e+01 &   97.742040 &  1.000000 &  0.000000e+00 \\
84         &   6.991360 &  3.884061e+01 &   19.634595 &  0.997704 &  4.843408e-07 \\
115        &  12.531356 &  3.360911e-08 &    0.000000 &  0.365079 &  3.992169e-02 \\
121        &   6.372688 &  1.603767e+00 &    0.000000 &  0.980718 &  3.204930e-03 \\
123        &  11.206271 &  3.784310e+01 &   30.296372 &  0.998778 &  3.147883e-02 \\
160        &   9.775328 &  7.847438e+01 &   41.454500 &  0.970033 &  1.562465e-01 \\
\bottomrule
\end{tabular}
\end{table}

\begin{table}[]
\centering
\caption{OLS regression model between Huff model performance and indicators. gas station (5541)}
\label{table:regression_5541}
\begin{tabular}{ccc}\hline
Indicator                 & $\beta$ coefficient & Confidence interval (95\%)   \\ \hline\hline
Mobility diversity        & $-0.3949$            & $[-1.2822, 0.4924]$  \\
Merchant diversity        & $0.0228$             & $[-1.0236, 1.0692]$  \\
Merchant monopoly         & $0.2326$             & $[-0.5643, 1.0295]$  \\
Gender diversity          & $-0.0024$            & $[-2.4916, 2.4868]$   \\
Marital status diversity  & $0.1781$             & $[-2.0838, 2.4400]$ \\
Education level diversity & $-0.5003$            & $[-1.4600, 0.4593]$ \\
Job status diversity      & $0.2611$             & $[-0.5382, 1.0604]$  \\
Income inequality         & $0.0928$             & $[-0.9343, 1.1200]$  \\ \hline
\end{tabular}
\end{table}

\begin{table}[]
\centering
\caption{OLS regression model between Huff model performance and indicators. Restaurants (5812)}
\label{table:regression_5812}
\begin{tabular}{ccc}\hline
Indicator                 & $\beta$ coefficient & Confidence interval (95\%)   \\ \hline\hline
Mobility diversity        & $0.2624$            & $[-0.4601, 0.9849]$  \\
Merchant diversity        & $-0.2362$           & $[-1.0883, 0.6159]$  \\
Merchant monopoly         & $-0.0775$           & $[-0.7264, 0.5715]$  \\
Gender diversity          & $-0.6746$           & $[-2.7015, 1.3524]$   \\
Marital status diversity  & $0.1780$            & $[-1.6638, 2.0199]$ \\
Education level diversity & $0.2504$            & $[-0.5310, 1.0319]$ \\
Job status diversity      & $-0.1828$           & $[-0.8337, 0.4680]$  \\
Income inequality         & $-0.4394$           & $[-1.2758, 0.3970]$  \\ \hline
\end{tabular}
\end{table}

\bibliographystyle{spbasic}
\bibliography{huff-model-paper}

\end{document}